\documentclass[conference]{IEEEtran}
\usepackage{textgreek}
\usepackage{blindtext}
\usepackage{graphicx}
\usepackage{color}
\usepackage{gensymb}
\usepackage{array}
\newcolumntype{P}[1]{>{\centering\arraybackslash}p{#1}}
\ifCLASSINFOpdf
\else
\fi
\begin{document}
\title{No Peeking through My Windows: Conserving Privacy in Personal Drones}

\author{
\IEEEauthorblockN{Alem Fitwi$^{1}$, Yu Chen$^{1}$, Sencun Zhu$^{2}$}
\IEEEauthorblockA{$^{1}$Dept. of Electrical and Computer Engineering,
Binghamton University, SUNY,  Binghamton, NY 13902, USA \\
$^{2}$Department of Computer Science and Engineering, Penn State University, University Park, PA 16802, USA\\
Emails: \{afitwi1, ychen\}@binghamton.edu, sxz16@psu.edu}
}
\maketitle
\begin{abstract} 
The drone technology has been increasingly used by many tech-savvy consumers, a number of defense companies, hobbyists and enthusiasts during the last ten years. Drones often come in various sizes and are designed for a multitude of purposes. Nowadays many people have small-sized personal drones for entertainment, filming, or transporting items from one place to another. However, personal drones lack a privacy-preserving mechanism. While in mission, drones often trespass into the personal territories of other people and capture photos or videos through windows without their knowledge and consent. They may also capture video or pictures of people walking, sitting, or doing private things within the drones' reach in clear form without their go permission. This could potentially invade people's personal privacy. This paper, therefore, proposes a lightweight privacy-preserving-by-design method that prevents drones from peeking through windows of houses and capturing people doing private things at home. It is a fast window object detection and scrambling technology built based on image enhancing, morphological transformation, segmentation and contouring processes (MASP). Besides, a chaotic scrambling technique is incorporated into it for privacy purpose. Hence, this mechanism detects window objects in every image or frame of a real-time video and masks them chaotically to protect the privacy of people. The experimental results validated that the proposed MASP method is lightweight and suitable to be employed in drones, considered as edge devices.
\end{abstract}

\IEEEpeerreviewmaketitle

\section{Introduction}
\label{sec:intro}
The world has seen a rampant advancement of unmanned aerial vehicles (UAV) technology, also known as drones, over the last decade. They come in a variety of size, sophistication, and they take many roles. Today, they have a wider range of applications in transportation, search and rescue, military, surveillance, communication relays, filming, entertainment, and monitoring \cite{fitwi2019agent, gharibi2016internet, roder2018unmanned, wang2018bandwidth,zeng2016wireless}. As a result, the public and private sectors, and individuals have shown growing interest in using the drone technologies in a way that serves their purposes. This is likely to cause proliferation of drones in the sky and these drones are capable of garnering a lot of private information about people and places \cite{cavoukian2012privacy}. While hovering in the sky, drones can be directed to collect personal information, monitor and spy people. 

Furthermore, this information can be divulged into the wider cyberspace because drones are vulnerable to a range of attacks. The owners might not have full control of their drones. Drones are vulnerable to rudimentary interception and interruption attacks. A number of attacks on drones like video stealing, injection of malwares, and device hijacking have been reported since 2007 \cite{arthur2009skygrabber, fitwi2019agent, javaid2012cyber}. It was clearly demonstrated that some commercially available WiFi based drones and satellite based military-grade ones are vulnerable to basic security attacks \cite{fifield2016drones, fitwi2019agent, hooper2016securing}. This has the potential to cause the widespread of sensitive personal data garnered by the drones without the owners' permission into the cyberspace. For this reason, people have been growing more and more paranoid about their privacy in relation to the use of drones. To some people, it feels like drones and the invasion of privacy are synonyms. 

In reality, personal or civilian drones have the capability to pick up virtually every information about what is happening in a certain specific scene. They can capture the images and record the footage of people in that scene without having their sanction to record them. The privacy of people is therefor at risk due to the fact that information, video or images unauthorizedly captured by drones' cameras and sensors could be abused by the owner of the drone and attackers who manage to penetrate into the drones \cite{fifield2016drones, mckelvey2019drones, rao2016societal, vattapparamban2016drones}. As a result, there have been a number of moves to enact laws and regulations to protect privacy and ensure safety in a bunch of developed countries where drones are widely used. For instance, UK, USA, and Canada have recently tried to legislate some laws and regulations in an attempt to address the privacy issues and to ensure safety in the aftermath of some incidents \cite{koerner2014drones, mcneal2014drones, schlag2012new, stanley2011protecting, takahashi2012drones}. However, this is not enough to preserve privacy. In fact, they have failed to address the burning privacy concerns.

This paper introduces a privacy-conserving technique for personal drones to balance out their use and privacy. A privacy-preserving mechanism based on germane image processing technique is proposed where the personal drones are considered as edge devices with single board computers (SBC) like Raspberry PI 3 B\textsuperscript{+}. It detects window objects in images or real-time video frames and automatically scrambles the windows to prevent peeking through them in violation of the privacy rights of people inside the house. It focuses on enabling the construction of privacy-aware drone systems by design. The window-object-detection and scrambling method are proposed and designed based on a less resource-intensive and faster morphological and segmentation process (MASP) that exploits the very nature of windows. The scrambling method incorporated as part and parcel of the MASP is designed based on a random chaos. Hence, our proposed MASP system is capable of detecting and scrambling window objects in every image or real-time video frames captured by drones.

The remainder of this paper is then organized as follows. Section \ref{sec:rel} presents background knowledge and previous work related to this paper. The window objects detection through morphological and segmentation techniques is introduced in Section \ref{sec:MAS}. Section \ref{sec:Exp}  reports the experimental results. The conclusions and future works are then presented in Section \ref{sec:con}.
\section{Background Information And Related Works}
\label{sec:rel}
Drones' popularity has increased beyond measures and they are employed in a number of areas including filming, relaying wireless connections, monitoring, search and rescue, military, transportation, and surveillance \cite{fitwi2019agent,gharibi2016internet,hooper2016securing, ref3, zeng2016wireless}. But most of them are not designed and manufactured with serious consideration of privacy and security. They are prone to basic attacks and the information they collect and carry can be compromised. Researchers have showed that personal drones are vulnerable to cache-poisoning and buffer overflow that have the potential of causing Denial of Service (DoS) attacks. The penetration tests that were conducted on personal drones like the Wireless Parrot Bebop UAVs revealed the aforementioned vulnerabilities \cite{gharibi2016internet, hooper2016securing, javaid2012cyber}. Some reported cases like the interception of live feeds from US drones by Iraqi militants who managed to access and watch captured videos on their laptops using a \$26 worth software shows that even military-grade drones are vulnerable \cite{cuadra2014drones, rao2016societal}. Hence, the information garnered and carried by cameras and sensors on personal drones operated by hobbyists and enthusiasts can be intercepted by attackers. That is, security problems of drones increases the degree of breaches of privacy of people whose personal details have been captured by such drones. The personal data might be disseminated into the vast cyberspace and become accessible to many wrong hands on top of the owners of the drones. Consequently, there have been ongoing efforts to bring about both legislative and technological solutions to the serious privacy issues in relation to the prolifically growing use of drones.

In parallel to the regulatory moves, endeavors have been underway to address the privacy issues of drones technologically by design. Efforts were exerted to develop security and privacy aware frameworks for drones \cite{javaid2012cyber, gharibi2016internet,ref3,zeng2016wireless}. Several security challenges that endanger the privacy of drones were studied and analyzed, which led to the proposal of cyber-security threat models. The models play very pronounced roles in aiding both users and designers of UAVs in understanding the possible threats with ease, vital for implementing some solutions. They, however, lack the ways and means for ensuring security and preserving privacy. They are only underlying frameworks on which one can build security and privacy solutions.  In addition, there are attempts to leverage the advancement in machine learning technology, chaotic cryptography and image scrambling techniques to conserve privacy in drones and surveillance systems \cite{al2019privacy, chaudhuri2009privacy, streiffer2017eprivateeye, wang2018enabling, yu2017iprivacy}. These works demonstrate how to build privacy-aware real-time video analytics in the surveillance systems; they are compute-intensive and impractical to be deployed on the edge, though. There are also many video and image scrambling methods \cite{dufaux2011video, fitwi2011performance, rahman2010real, zhang2013chaos}. But they need to be redesigned for drones where resources are constrained. Besides, the Morphological techniques \cite{serra2012mathematical, salembier1994hierarchical} could be vital in processing images containing regularly shaped objects.

\section{MASP: Windows Detection and Scrambling}
\label{sec:MAS}

By their very nature, windows comprise horizontal and vertical edges, and some of them contains arcs. They could also be multi-framed comprising multiple vertical and horizontal edges and lines. The edges are represented by changes from one class of pixels to another, and basically refers to the boundaries of windows. Whereas the lines refer to one class of pixels interlaced between two sets of the same class of pixels. They are the metal or wooden lines that partition a window into multiple panes.  Besides, in most cases they are rectangular or closely rectangular shaped or combined rectangular and semi-circular shapes. Hence, in this paper we introduce a window detection and scrambling technique that exploits these natures of windows, which is a faster and computing-resources conscious method for resource-limited environments like drones. It is designed based on low-level, medium-level, and high-level image manipulations and processing techniques. The major steps include image enhancing, morphological processing, semantic segmentation, recognition, and scrambling.
\subsection{Image/Video Frames Pre-processing}
\label{sec:img} 
\begin{figure}[t]
    \centering
        \includegraphics[width=0.4\textwidth]{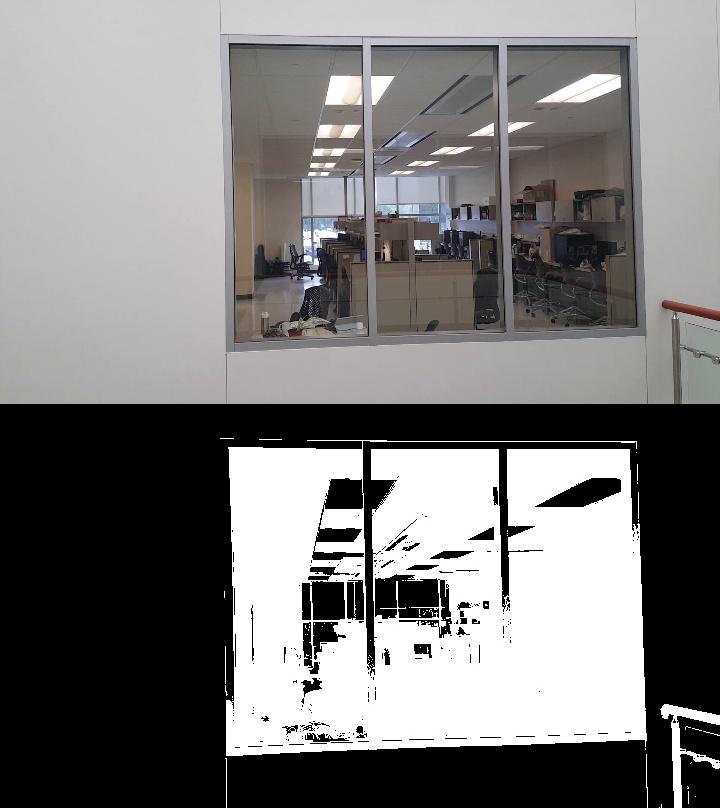}
    \caption{Image/Frame pre-processing, f(x,y)\textsubscript{input} vs f(x,y)\textsubscript{INV}}
    \label{fig_3}
\end{figure}

Following the acquisition of images as they are and videos split into frames, they are converted into gray-scale pictures in order to reduce the processing complexity. The assumption here is that images or video frames contain windows and walls. Hence, they are bi-modal in that they contain two classes of pixels, namely the window (foreground) pixels and the wall (background) pixels. A method  of thresholding on the gray-scale image/frame \cite{otsu1979threshold} is adopted to compute the optimum threshold that separates the two classes. It exhaustively searches for the threshold that cuts down on the intra-class variance defined by Eq. (\ref{eq:1}) where variances of each class is multiplied by the class weight.
   \begin{equation}
    \label{eq:1}
    \delta_{overall}^{2}(t)=w_{win}(t)\delta_{win}^{2}(t)+w_{wall}(t)\delta_{wall}^{2}(t)
  \end{equation}
   \begin{equation}
    \label{eq:2}
    f(x,y)_{INV}=f(x,y)_{gray}-255
  \end{equation}
Then, inverting the gray image (f(x,y)\textsubscript{gray}) using Eq. (\ref{eq:2}) after the binarization has been applied gives a white window and a black wall or background as portrayed in Fig. \ref{fig_3}. The inversion is done by subtracting 255 from every pixel.
\subsection{Morphological Transformations}
\label{sec:mor} 
\begin{figure}[t]
    \centering
        \includegraphics[width=0.4\textwidth]{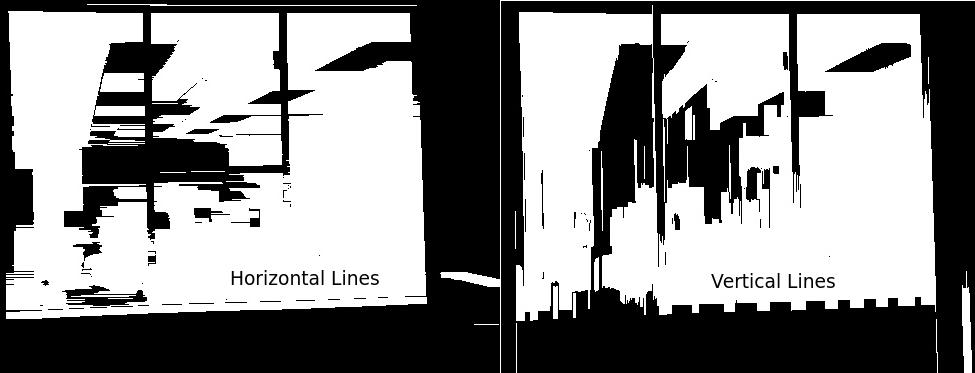}
    \caption{Horizontal versus vertical lines.}
    \label{fig_4}
    \vspace{-10pt}
\end{figure}

Then, features important for detecting the window object are extracted from the inverted image using a morphological process performed based on the shape of target objects in the image. It works with a kernel that slides around to eliminate noises. That is, we employed rectangular morphology for detecting the horizontal and vertical lines and edges useful to detect the windows in the image because most windows have approximately rectangular shapes. The window object dimensions are assumed to be smaller than that of the full image. In most energy efficient buildings' design, the maximum window-to-wall ratio (WWR) is 60\%. Hence, basically we adopted 3x3 kernel whose horizontal and vertical lengths are (image\_width-8) and (image\_height-8), respectively. These selections were made following a string of tests and consultations to building standards. Figure \ref{fig_4} portrays the output of the morphological transformation which combines the images of detected horizontal and vertical lines based on two weights, each assigned a value of 0.5.
\subsection{Semantic Segmentation}
\label{sec:seg} 
Pre-processed images are then divided into segments that contain meaningful objects, window objects in this case. That is, the segmentation process groups together the pixels that have similar attributes. For we are working on bi-modal images where there are two classes, windows and walls, a semantic segmentation is employed. The pixels that belong to the window object are represented by one color and those pixels that belong to the wall or background class are represented by another class of color.
\begin{figure}[t]
    \centering
        \includegraphics[width=0.4\textwidth]{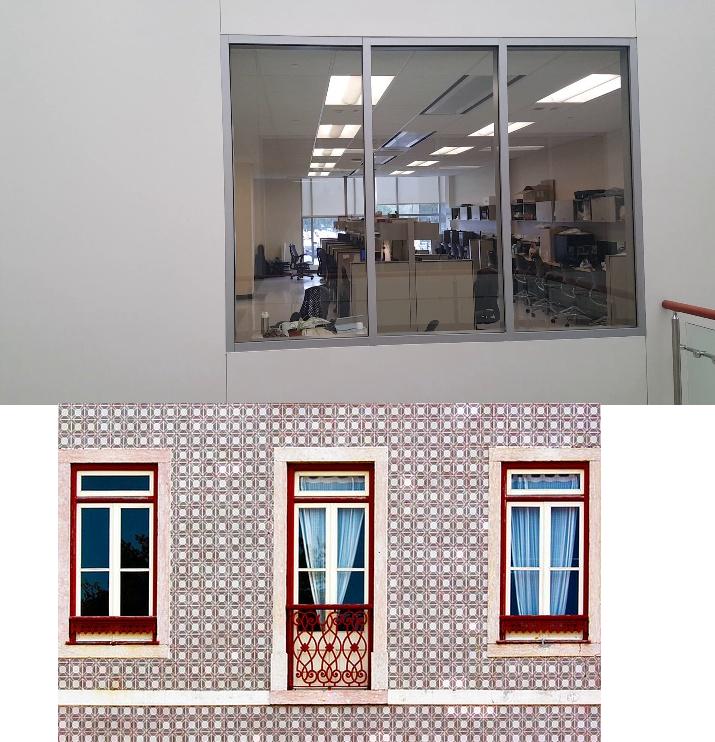}
    \caption{Two input images to the system [top and bottom].}
    \label{fig_5}
\end{figure}
\begin{figure}[t]
    \centering
        \includegraphics[width=0.4\textwidth]{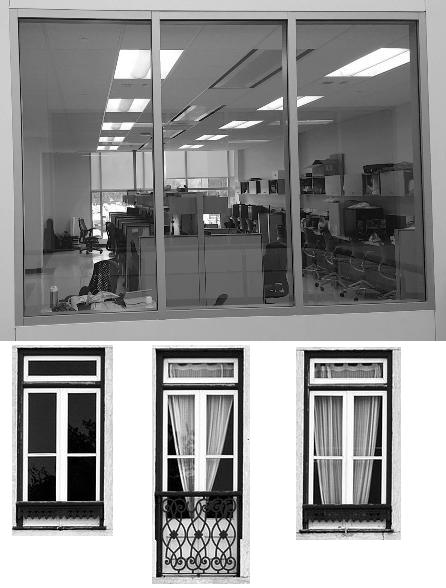}
    \caption{Segmented and Extracted Windows.}
    \label{fig_6}
    \vspace{-10pt}
\end{figure}

Following the segmentation process, we used a contouring technique based on the Suzuki algorithm \cite{suzuki1985topological} to extract the location coordinates of all window objects on every input image/video frame. Figure \ref{fig_5}  gives two input images: the one on top containing only a window and the other (on the bottom) containing three window objects. Figure \ref{fig_6} illustrates the correctly extracted windows from the input images. 
\subsection{Window Objects Recognition and Scrambling}
\label{sec:recog} 
Now that the coordinates of the window objects in the frames are known, they can be easily located and recognized. The next step is scrambling all windows to eschew any look-through. The scrambling process could be done using simple image denaturing processes but we have opted to employ a more robust and reversible chaotic scrambling. 

The scrambling is then done using chaotic images generated by solving a system of differential equations stated in Eq. (\ref{eq:3}) and Eq.  (\ref{eq:4}) \cite{fitwi2011performance}.
   \begin{equation}
    \label{eq:3}
    \frac{dy}{dt^{2}}=2\alpha\frac{dx}{dt}-x 
  \end{equation}
   \begin{equation}
    \label{eq:4}
    \frac{d^{2}x}{dt^{2}}-2\alpha\frac{dx}{dt}+x=0 
  \end{equation}
  
Eq. (\ref{eq:4}) is a homogeneous equation of the form  described in Eq. (\ref{eq:5}) where $f(x)=0$.
   \begin{equation}
    \label{eq:5}
    a\frac{d^{2}x}{dt^{2}}+b\frac{dx}{dt}+cx=f(x) 
  \end{equation}
It produces chaos only if its solutions are complex. Then, solving Eq. (\ref{eq:4}) using this requirement and simplifying it using Euler's formula, it produces the system of solutions stated in Eqs. (\ref{eq:6}) and (\ref{eq:7}).
   \begin{equation}
    \label{eq:6}
    x(t)=e^{\alpha{t}}[x_{0}cos(\beta{t})+(\frac{y_{0}-\alpha{x_{0}}}{\beta})sin(\beta{t})]
  \end{equation}
   \begin{equation}
    \label{eq:7}
    y(t)=e^{\alpha{t}}[y_{0}cos(\beta{t})+\frac{\alpha}{\beta}(y_{0}-\alpha{x_{0}}-\frac{\beta^{2}}{\alpha}x_{0})sin(\beta{t})]
  \end{equation}
  
\noindent where $x_{0}$ and $y_{0}$ are initial conditions, and $\alpha$ and $\beta$ are constants related to each other by Eq. (\ref{eq:8}): 
  
   \begin{equation}
    \label{eq:8}
    \beta=\sqrt{1-\alpha^{2}}
  \end{equation}

The $\alpha$ and $\beta$ values must be carefully selected so as to produce a complex solution and then a random enough chaos. After researching on a range of values, we eventually chose $\alpha=0.005$, and then $\beta$ was computed to be 0.999 using Eq. (\ref{eq:8}). With these values, we were able to produce a randomized chaos. The good thing about a chaotic scrambling is that it could be reproduced using the same initial conditions ($x_{0}$ and $y_{0}$) and constant values ($\alpha$ and $\beta$) whenever deemed necessary. The chaotic generator is more like the pseudo-random generators where the initial conditions and constants serve as the seed value.
 
Now the privacy-conserving mechanism is ready to detect window objects on any image or video frame by means of morphological transformation, semantic segmentation, and the Suzuki contouring algorithm. Then, it performs the scrambling of window objects through chaotic mixing for privacy reasons. The chaotic scrambling technique is embedded into the privacy-preserving mechanism and it automatically scrambles once a window object has been detected on the input frames. 
\section{Experimental Settings and Results}
\label{sec:Exp}
\subsection{Experimental Setup}
\label{sec:TTI}
At last, we carried out a number of experiments and tests treating the drones as edge devices with constrained resources to verify our proposed model. Then, in terms of the experimental setup, a single board computer (SBC) is employed. It is a Raspberry PI 3 Model B\textsuperscript{+} with 1.4GHz Cortex-A53 (ARMv8) 64-bit quad core CPU and 1GB DDR2 RAM. It is also fitted with a camera for real-time video analytics testing. Besides, we run our model on an old laptop with Intel(R) Core(TM) i5-3337U CPU @ 1.80GHz and 3938MB System memory in an effort to observe how the performance varies with respect to computing device changes. The testing was conducted on a Raspberry Pi 3 B\textsuperscript{+} connected to a live camera to produce the result on Fig. \ref{fig_8}.

\subsection{Results and Discussion}
\label{sec:RaD}
 
\begin{figure}[t]
    \centering
        \includegraphics[width=0.4\textwidth]{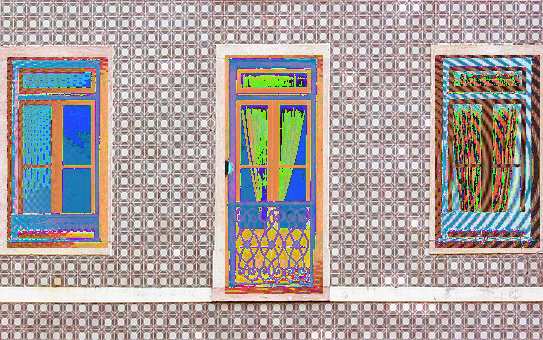}
    \caption{Sample image test result.}
    \label{fig_7}
    \vspace{-10pt}
\end{figure}

\begin{figure}[t]
    \centering
        \includegraphics[width=0.4\textwidth]{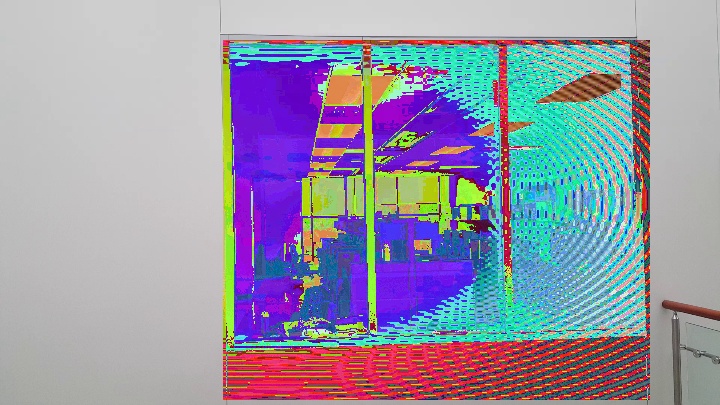}
    \caption{Sample real-time video test result.}
    \label{fig_8}
    \vspace{-10pt}
\end{figure}

 To verify the functionality, speed, and accuracy of our proposed model, we conducted extensive tests using 17,400 images, recorded and real-time videos of buildings containing various types of windows. Out of the 17,400 images, 11,600 of them are positive images that contain windows of different style, color, size, and varying positions. The rest 5,800 images are dummy walls or negative images that don't contain windows. For demonstration purpose, the video frame and image in Fig. \ref{fig_5} are used. The test image (bottom) is one of the many different types of windows used and the video is one of the many different videos recorded at different scenes and tested live in real-time. As can be seen from the sample test image result in Fig. \ref{fig_7}, all of the three windows were effectively detected and scrambled. Likewise, all window objects in every frame of a video captured in a scene in our campus were successfully detected and scrambled. The sample result is portrayed in Fig. \ref{fig_8}. Our method is primarily designed to make window objects detection and automatic scrambling in real-time video analytics. 
 
\begin{figure}[t]
    \centering
        \includegraphics[width=0.4\textwidth]{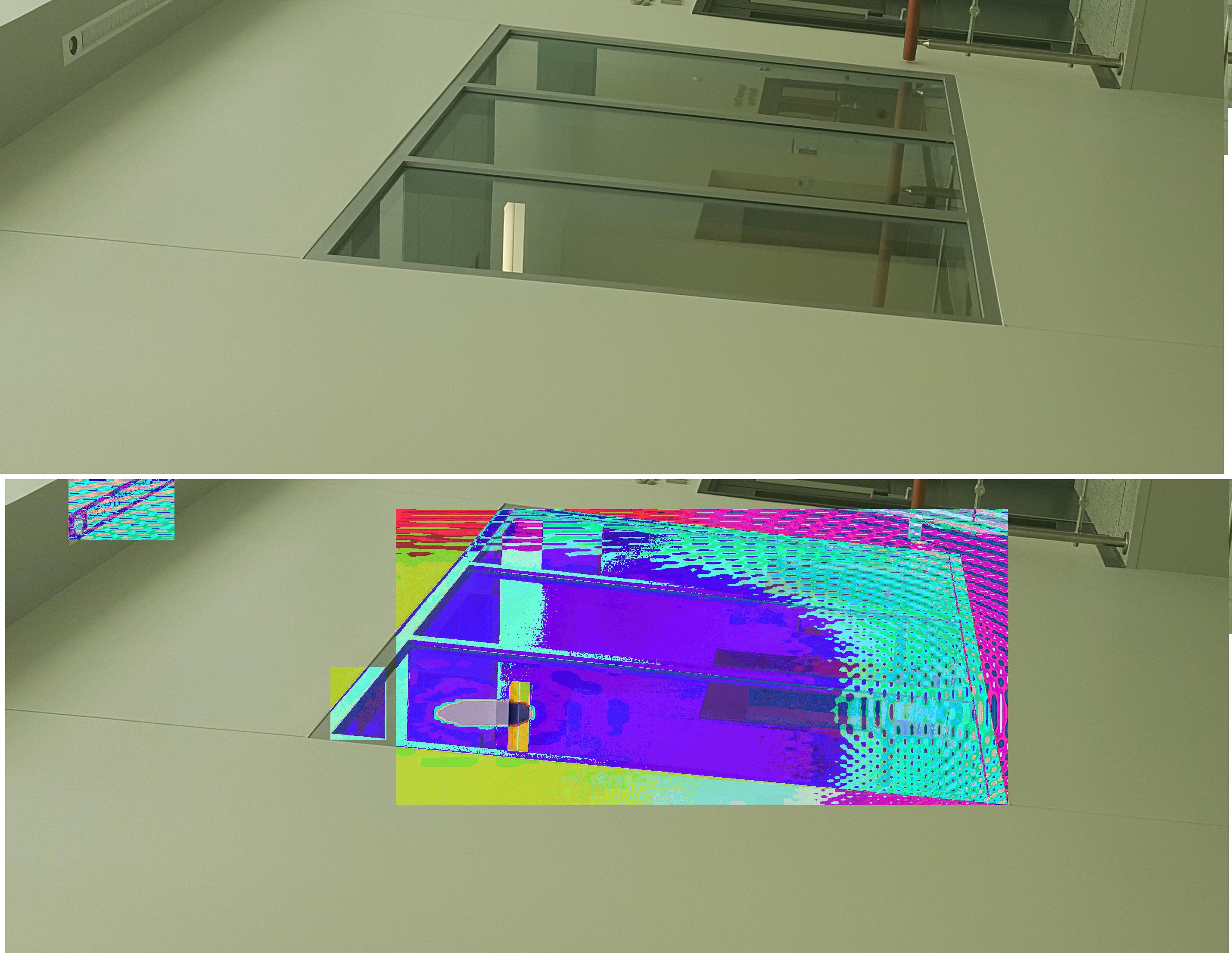}
    \caption{A window object captured at slant angle and scrambled multiple times.}
    \label{fig_17}
    \vspace{-10pt}
\end{figure}

In the MASP scheme, a window object is detected based on basic image processing, morphological transformation, semantic segmentation, and contouring process. The results are very promising. It can process more than 30 frames per second (FPS) on average on the old laptop described in section \ref{sec:TTI} and more than 8 FPS on the Raspberry PI 3 B\textsuperscript{+} as portrayed in table \ref{tab:t1}. The weights used in our experiments for the window and background image classes are $w_{win}=0.5$ and $w_{bgd}=0.5$, respectively. The only shortcoming observed is that it considers a window object in an image or video frame captured from very slant angles as multiple windows. We found out that a window object captured at angle greater than $60^{o}$ clockwise or counter-clockwise from the normal line to the window surface is treated as if comprising multiple windows and perform the scrambling multiple times as portrayed in Fig. \ref{fig_17}. Small window-like non-window shapes are also detected as windows and scrambled as illustrated in Fig. \ref{fig_17}. But it is still a very decent method because when the distance from which such windows are shot becomes longer, the small window-like shapes are less likely to be detected. In case of drones, the minimum distance as required by law from any property is 50 meters. We also measured the accuracy of the window objects detection based on the tests carried out on the 17,400 images and a number of real-time videos. It detects windows 100\% at 8.69 FPS but we found out that it has an average false positive rate (FPR) of 10.67\%. 

\begin{table}[ht]
\centering
\caption{Number of FPS required for Laptop and Raspberry PI}
\label{tab:t1}
\begin{center}
\begin{tabular}{|l|p{0.9cm}|p{2.5cm}|p{1cm}|}
\hline
\rule[-1ex]{0pt}{3.5ex} \textbf{Machine} & \textbf{Laptop} & \textbf{Raspberry PI 3 B\textsuperscript{+}}\\
\hline\hline
\rule[-1ex]{0pt}{3.5ex} \textbf{FPS} & \textit{33.25} & \textit{8.69}\\
\hline\hline
\rule[-1ex]{0pt}{3.5ex} \textbf{FPR} & N/A & \textit{10.67\%}\\
\hline
\end{tabular}
\end{center}
\end{table}
 For comparative analysis, we have researched on preexisting algorithms including histogram oriented gradients (HOG) features based methods like Support Vector Machine (SVM) classifier coupled with HOG features \cite{pang2011efficient},  Convolutional neural networks (CNN) based methods like Faster Region-based CNN (Faster R-CNN) \cite{ren2015faster} and Spatial Pyramid Pooling(SPP-net) \cite{he2015spatial}, regression-based object detectors like You only Look Once version 3 (YOLOv3)\cite{redmon2018yolov3} and the Single Shot Detector(SSD) \cite{liu2016ssd}, and the the Haar-cascade network \cite{xu2016enhanced}. They are very promising methods; however, they are compute-intensive in that they cannot run well at the edge,  a resource-constrained environment. 
 
 We discovered that most of these methods can process images or video frames at less than 2 frame per second (fps) on a Raspberry Pi 3 B\textsuperscript{+}. A benchmark made on how much time each of the models can take to make a prediction on a new image shows that the fastest method is the SDD\_mobilenet\_v1 which can process 1.39 frames per second or it takes 0.72 seconds for processing an image on the Raspberry Pi. However, there is a more recently (in 2018) developed light-weight CNN-based object detector \cite{nikouei2018smart} where the fps is about 1.79 when run on an edge device. Harkening back to our experiments, we trained the HOG+SVM, and Haar-Cascade networks using the dataset we created and found out the results portrayed in table \ref{tab:t2}. The results substantiate that our proposed model is much faster on the edge than any object-detection method made available in the public domain of the Internet. It is computing resource-conscious method of window-object-detection and scrambling. But there is a room for improving its FPR.
 
\begin{table}[ht]
\centering
\caption{Performance Measures in terms of FPR and FPS.}
\label{tab:t2}
\begin{center}
\begin{tabular}{|l|p{2.5cm}|p{2.5cm}|p{1cm}|}
\hline
\rule[-1ex]{0pt}{3.5ex} \textbf{Methods} & \textbf{False Positives Rate} & \textbf{Frames Per Second}\\
\hline\hline
\rule[-1ex]{0pt}{3.5ex} \textbf{MASP} & \textit{10.67\%} & \textit{8.69}\\
\hline\hline
\rule[-1ex]{0pt}{3.5ex} \textbf{HOG+SVM} & \textit{13.23\%} & \textit{0.65}\\
\hline\hline
\rule[-1ex]{0pt}{3.5ex} \textbf{Haar} & \textit{22.71\%} & \textit{1.93}\\
\hline\hline
\end{tabular}
\end{center}
\end{table}

In summary, our experimental results corroborate the feasibility and suitability of our proposed model for drones, considered as as edge devices. It is capable of processing a decent FPS on a resource-constrained environment, where there are only a CPU of 1.4GHz and a memory of 1GB. 
\section{Conclusions and Future Works}
\label{sec:con}

Passing laws and imposing regulations are vital for safe operation of drones and for creating non-flying sky zones. However, they cannot efficiently address the invasion and violation of the privacy rights of individuals in relation to the prolific use of drones. They give the slightest concern to privacy. The best solution is, therefore, to incorporate a privacy-conserving mechanism into the drones by design. This has the potential to play very pronounced role in solving the issues of privacy without impeding the technological advancement of drones. Thus, this paper proposes a method to address the the privacy issues by design. 
 
The method we proposed for conserving privacy in personal drones take advantages of the nature of windows, image processing, and simple edge detection techniques. As a result, it is faster. The light-weight MASP can process more than 8 fps on a resource-constrained Raspberry PI 3 B\textsuperscript{+}. 

The contents reported in this paper are results of our preliminary study. It does not handle windows with stylish shapes, i.e. with arc, or being embedded in a more complex walls or houses. Also, it suffers from a relatively high false positive rate (10.67\%). Being aware of these weaknesses, our on-going efforts are exploring a machine learning approach that leverages the powerful convolutional neural networks to improve the detection accuracy in more complex scenarios and a great deal of windows with different styles. For this purpose, a focus is being given to the design of a lightweight machine learning algorithm that can tackle the resource constraints on the drones. Beyond windows, the detection and denaturing of human faces or other sensitive parts of the human body are investigated to address other privacy concerns, such as personal identities, children protection, etc.

\ifCLASSOPTIONcaptionsoff
  \newpage
\fi
\bibliographystyle{IEEEtranS}
\bibliography{Ref.bib}

\begin{thebibliography}{10}
\providecommand{\url}[1]{#1}
\csname url@samestyle\endcsname
\providecommand{\newblock}{\relax}
\providecommand{\bibinfo}[2]{#2}
\providecommand{\BIBentrySTDinterwordspacing}{\spaceskip=0pt\relax}
\providecommand{\BIBentryALTinterwordstretchfactor}{4}
\providecommand{\BIBentryALTinterwordspacing}{\spaceskip=\fontdimen2\font plus
\BIBentryALTinterwordstretchfactor\fontdimen3\font minus
  \fontdimen4\font\relax}
\providecommand{\BIBforeignlanguage}[2]{{%
\expandafter\ifx\csname l@#1\endcsname\relax
\typeout{** WARNING: IEEEtranS.bst: No hyphenation pattern has been}%
\typeout{** loaded for the language `#1'. Using the pattern for}%
\typeout{** the default language instead.}%
\else
\language=\csname l@#1\endcsname
\fi
#2}}
\providecommand{\BIBdecl}{\relax}
\BIBdecl

\bibitem{al2019privacy}
M.~Al-Rubaie and J.~M. Chang, ``Privacy-preserving machine learning: Threats
  and solutions,'' \emph{IEEE Security \& Privacy}, vol.~17, no.~2, pp. 49--58,
  2019.

\bibitem{arthur2009skygrabber}
C.~Arthur, ``Skygrabber: the \$26 software used by insurgents to hack into us
  drones,'' \emph{The Guardian}, vol.~17, 2009.

\bibitem{cavoukian2012privacy}
A.~Cavoukian, \emph{Privacy and drones: Unmanned aerial vehicles}.\hskip 1em
  plus 0.5em minus 0.4em\relax Information and Privacy Commissioner of Ontario,
  Canada Ontario, 2012.

\bibitem{chaudhuri2009privacy}
K.~Chaudhuri and C.~Monteleoni, ``Privacy-preserving logistic regression,'' in
  \emph{Advances in neural information processing systems}, 2009, pp. 289--296.

\bibitem{cuadra2014drones}
A.~Cuadra and C.~Whitlock, ``How drones are controlled,'' \emph{The Washington
  Post}, vol.~20, 2014.

\bibitem{dufaux2011video}
F.~Dufaux, ``Video scrambling for privacy protection in video surveillance:
  recent results and validation framework,'' in \emph{Mobile Multimedia/Image
  Processing, Security, and Applications 2011}, vol. 8063.\hskip 1em plus 0.5em
  minus 0.4em\relax International Society for Optics and Photonics, 2011, p.
  806302.

\bibitem{fifield2016drones}
J.~Fifield, ``How drones raised privacy concerns across cyberspace,'' 2016.

\bibitem{fitwi2019agent}
A.~Fitwi, Y.~Chen, and N.~Zhou, ``An agent-administrator-based security
  mechanism for distributed sensors and drones for smart grid monitoring,'' in
  \emph{Signal Processing, Sensor/Information Fusion, and Target Recognition
  XXVIII}, vol. 11018.\hskip 1em plus 0.5em minus 0.4em\relax International
  Society for Optics and Photonics, 2019, p. 110180L.

\bibitem{fitwi2011performance}
A.~H. Fitwi and S.~Nouh, ``Performance analysis of chaotic encryption using a
  shared image as a key,'' \emph{Zede Journal}, vol.~28, pp. 17--29, 2011.

\bibitem{gharibi2016internet}
M.~Gharibi, R.~Boutaba, and S.~L. Waslander, ``Internet of drones,'' \emph{IEEE
  Access}, vol.~4, pp. 1148--1162, 2016.

\bibitem{he2015spatial}
K.~He, X.~Zhang, S.~Ren, and J.~Sun, ``Spatial pyramid pooling in deep
  convolutional networks for visual recognition,'' \emph{IEEE transactions on
  pattern analysis and machine intelligence}, vol.~37, no.~9, pp. 1904--1916,
  2015.

\bibitem{hooper2016securing}
M.~Hooper, Y.~Tian, R.~Zhou, B.~Cao, A.~P. Lauf, L.~Watkins, W.~H. Robinson,
  and W.~Alexis, ``Securing commercial wifi-based uavs from common security
  attacks,'' in \emph{MILCOM 2016-2016 IEEE Military Communications
  Conference}.\hskip 1em plus 0.5em minus 0.4em\relax IEEE, 2016, pp.
  1213--1218.

\bibitem{javaid2012cyber}
A.~Y. Javaid, W.~Sun, V.~K. Devabhaktuni, and M.~Alam, ``Cyber security threat
  analysis and modeling of an unmanned aerial vehicle system,'' in \emph{2012
  IEEE Conference on Technologies for Homeland Security (HST)}.\hskip 1em plus
  0.5em minus 0.4em\relax IEEE, 2012, pp. 585--590.

\bibitem{ref3}
R.~Kinge, P.~Gawande, A.~S.~ingle, and S.~Badhe, ``Internet of drones,''
  \emph{nternational Journal of Research in Advent Technology (IJRAT) (E-ISSN:
  2321-9637)}, 2017.

\bibitem{koerner2014drones}
M.~R. Koerner, ``Drones and the fourth amendment: Redefining expectations of
  privacy,'' \emph{Duke LJ}, vol.~64, p. 1129, 2014.

\bibitem{liu2016ssd}
W.~Liu, D.~Anguelov, D.~Erhan, C.~Szegedy, S.~Reed, C.-Y. Fu, and A.~C. Berg,
  ``Ssd: Single shot multibox detector,'' in \emph{European conference on
  computer vision}.\hskip 1em plus 0.5em minus 0.4em\relax Springer, 2016, pp.
  21--37.

\bibitem{mckelvey2019drones}
N.~McKelvey, C.~Diver, and K.~Curran, ``Drones and privacy,'' in \emph{Unmanned
  Aerial Vehicles: Breakthroughs in Research and Practice}.\hskip 1em plus
  0.5em minus 0.4em\relax IGI Global, 2019, pp. 540--554.

\bibitem{mcneal2014drones}
G.~S. McNeal, ``Drones and aerial surveillance: Considerations for
  legislators,'' \emph{Brookings Institution: The Robots Are Coming: The
  Project on Civilian Robotics}, 2014.

\bibitem{nikouei2018smart}
S.~Y. Nikouei, Y.~Chen, S.~Song, R.~Xu, B.-Y. Choi, and T.~Faughnan, ``Smart
  surveillance as an edge network service: From harr-cascade, svm to a
  lightweight cnn,'' in \emph{2018 IEEE 4th International Conference on
  Collaboration and Internet Computing (CIC)}.\hskip 1em plus 0.5em minus
  0.4em\relax IEEE, 2018, pp. 256--265.

\bibitem{otsu1979threshold}
N.~Otsu, ``A threshold selection method from gray-level histograms,''
  \emph{IEEE transactions on systems, man, and cybernetics}, vol.~9, no.~1, pp.
  62--66, 1979.

\bibitem{pang2011efficient}
Y.~Pang, Y.~Yuan, X.~Li, and J.~Pan, ``Efficient hog human detection,''
  \emph{Signal Processing}, vol.~91, no.~4, pp. 773--781, 2011.

\bibitem{rahman2010real}
S.~M.~M. Rahman, M.~A. Hossain, H.~Mouftah, A.~El~Saddik, and E.~Okamoto, ``A
  real-time privacy-sensitive data hiding approach based on chaos
  cryptography,'' in \emph{2010 IEEE International Conference on Multimedia and
  Expo}.\hskip 1em plus 0.5em minus 0.4em\relax IEEE, 2010, pp. 72--77.

\bibitem{rao2016societal}
B.~Rao, A.~G. Gopi, and R.~Maione, ``The societal impact of commercial
  drones,'' \emph{Technology in Society}, vol.~45, pp. 83--90, 2016.

\bibitem{redmon2018yolov3}
J.~Redmon and A.~Farhadi, ``Yolov3: An incremental improvement,'' \emph{arXiv
  preprint arXiv:1804.02767}, 2018.

\bibitem{ren2015faster}
S.~Ren, K.~He, R.~Girshick, and J.~Sun, ``Faster r-cnn: Towards real-time
  object detection with region proposal networks,'' in \emph{Advances in neural
  information processing systems}, 2015, pp. 91--99.

\bibitem{roder2018unmanned}
A.~Roder, K.-K.~R. Choo, and N.-A. Le-Khac, ``Unmanned aerial vehicle forensic
  investigation process: Dji phantom 3 drone as a case study,'' \emph{arXiv
  preprint arXiv:1804.08649}, 2018.

\bibitem{salembier1994hierarchical}
P.~J. Salembier~Clairon and M.~Pard{\`a}s~Feliu, ``Hierarchical morphological
  segmentation for image sequence coding,'' \emph{IEEE Transactions on Image
  Processing}, vol.~3, no.~5, pp. 639--651, 1994.

\bibitem{schlag2012new}
C.~Schlag, ``The new privacy battle: How the expanding use of drones continues
  to erode our concept of privacy and privacy rights,'' \emph{Pitt. J. Tech. L.
  \& Pol'y}, vol.~13, p.~i, 2012.

\bibitem{serra2012mathematical}
J.~Serra and P.~Soille, \emph{Mathematical morphology and its applications to
  image processing}.\hskip 1em plus 0.5em minus 0.4em\relax Springer Science \&
  Business Media, 2012, vol.~2.

\bibitem{stanley2011protecting}
J.~Stanley, C.~Crump, and A.~Speech, \emph{Protecting Privacy From Aerial
  Surveillance}.\hskip 1em plus 0.5em minus 0.4em\relax American Civil
  Liberties Union.(December 2011), 2011, vol.~6, no.~6.

\bibitem{streiffer2017eprivateeye}
C.~Streiffer, A.~Srivastava, V.~Orlikowski, Y.~Velasco, V.~Martin, N.~Raval,
  A.~Machanavajjhala, and L.~P. Cox, ``eprivateeye: To the edge and beyond!''
  in \emph{Proceedings of the Second ACM/IEEE Symposium on Edge
  Computing}.\hskip 1em plus 0.5em minus 0.4em\relax ACM, 2017, p.~18.

\bibitem{suzuki1985topological}
S.~Suzuki \emph{et~al.}, ``Topological structural analysis of digitized binary
  images by border following,'' \emph{Computer vision, graphics, and image
  processing}, vol.~30, no.~1, pp. 32--46, 1985.

\bibitem{takahashi2012drones}
T.~Takahashi, ``Drones and privacy,'' 2012.

\bibitem{vattapparamban2016drones}
E.~Vattapparamban, {\.I}.~G{\"u}ven{\c{c}}, A.~{\.I}. Yurekli, K.~Akkaya, and
  S.~Ulua{\u{g}}a{\c{c}}, ``Drones for smart cities: Issues in cybersecurity,
  privacy, and public safety,'' in \emph{2016 International Wireless
  Communications and Mobile Computing Conference (IWCMC)}.\hskip 1em plus 0.5em
  minus 0.4em\relax IEEE, 2016, pp. 216--221.

\bibitem{wang2018enabling}
J.~Wang, B.~Amos, A.~Das, P.~Pillai, N.~Sadeh, and M.~Satyanarayanan,
  ``Enabling live video analytics with a scalable and privacy-aware
  framework,'' \emph{ACM Transactions on Multimedia Computing, Communications,
  and Applications (TOMM)}, vol.~14, no.~3s, p.~64, 2018.

\bibitem{wang2018bandwidth}
J.~Wang, Z.~Feng, Z.~Chen, S.~George, M.~Bala, P.~Pillai, S.-W. Yang, and
  M.~Satyanarayanan, ``Bandwidth-efficient live video analytics for drones via
  edge computing,'' in \emph{2018 IEEE/ACM Symposium on Edge Computing
  (SEC)}.\hskip 1em plus 0.5em minus 0.4em\relax IEEE, 2018, pp. 159--173.

\bibitem{xu2016enhanced}
Y.~Xu, G.~Yu, X.~Wu, Y.~Wang, and Y.~Ma, ``An enhanced viola-jones vehicle
  detection method from unmanned aerial vehicles imagery,'' \emph{IEEE
  Transactions on Intelligent Transportation Systems}, vol.~18, no.~7, pp.
  1845--1856, 2016.

\bibitem{yu2017iprivacy}
J.~Yu, B.~Zhang, Z.~Kuang, D.~Lin, and J.~Fan, ``iprivacy: image privacy
  protection by identifying sensitive objects via deep multi-task learning,''
  \emph{IEEE Transactions on Information Forensics and Security}, vol.~12,
  no.~5, pp. 1005--1016, 2017.

\bibitem{zeng2016wireless}
Y.~Zeng, R.~Zhang, and T.~J. Lim, ``Wireless communications with unmanned
  aerial vehicles: Opportunities and challenges,'' \emph{IEEE Communications
  Magazine}, vol.~54, no.~5, pp. 36--42, 2016.

\bibitem{zhang2013chaos}
X.~Zhang and X.~Wang, ``Chaos-based partial encryption of spiht coded color
  images,'' \emph{Signal Processing}, vol.~93, no.~9, pp. 2422--2431, 2013.

\end{thebibliography}
\end{document}